# Full-field fluorescence computed tomography (F3CT) using a calibrated virtual cone-beam pinhole geometry

Thomas Zillhardt[1,2*], Yunhui Chen[1,3,5], Alexander Rack[3], Matthew Veale[4], Matt Wilson[4], Philip J. Withers[1,6], Nicola Vigano[3,7]

[1] Henry Royce Institute, Department of Materials, University of Manchester, Oxford Road, M13 9PL, Manchester, United Kingdom

[2] Diamond Light Source, Diamond House: Harwell Science and Innovation Campus, Didcot, Oxfordshire, OX11 0DE, United Kingdom

[3] The ESRF - European Synchrotron, 71 Avenue des Martyrs, 38000 Grenoble, France

[4] STFC-UKRI, Rutherford Appleton Laboratory, Harwell, OX11 0QX, UK

[5] RMIT University, GPO Box 2476, Melbourne VIC 3001 Australia

[6] Department of Materials Science and Engineering, Monash University, Clayton, Vic-3800, Australia

[7] Université Grenoble Alpes, CEA, IRIG-MEM, Grenoble, 38000, France

---

Keywords: X-ray fluorescence tomography, Full-field imaging, Hyperspectral imaging, Pinhole tomography, Correlative tomography

## Abstract

We present F3CT, a synchrotron-based hyperspectral full-field fluorescence computed tomography technique that avoids raster scanning by combining a pinhole aperture with an energy-resolving 2D detector. A virtual cone-beam model and two-stage calibration–reconstruction workflow enable 3D elemental mapping under full-field illumination. The method is demonstrated on biological and geological specimens, resolving silver-stain distributions in zebrafish tissue and high-energy fluorescence signatures in rock cores. Phase-contrast tomograms acquired sequentially under the same experimental geometry provide co-registered structural context. F3CT provides a high-throughput route to 3D XRF imaging and establishes a foundation for correlated structural and chemical tomography, with potential for future in-situ and operando implementations.

# 1 Introduction

X-Ray Fluorescence (XRF) imaging provides chemically selective contrast in two and three dimensions across a wide range of length scales. It is widely used in materials science[1], chemistry[2], and cultural heritage[3] due to its high chemical sensitivity. This contrast arises from characteristic secondary photons emitted following core-level excitation, allowing energy-resolving detectors to identify elemental contributions within the sample[4,5]. Energy-resolving detectors can distinguish these photons, enabling precise identification of the elements in the sample. XRF imaging typically involves scanning samples with a monochromatic focused beam (or pencil beam) and collecting the emitted fluorescence signal with single-pixel energy-resolving detectors (EDS) or using diffraction gratings to separate the X-rays according to their wavelength (WDS). The collected spectra are then processed to determine the local chemical composition. Phase contrast imaging (PCI), on the other hand, provides high contrast and high-resolution morphological information, making it invaluable for detailed structural analysis.

Tomographic reconstruction enhances the capabilities of XRF and PCI by extending them to three-dimensional analysis. XRF computed tomography (XRF-CT) reconstructs the spatial distribution of elements within a sample, while phase contrast tomography (PCT) provides detailed three-dimensional morphological information. Volumetric insight is essential for understanding complex structures and dynamic processes that cannot be fully captured by 2D imaging. However, conventional XRF-CT is most commonly implemented via raster-scanning the sample with a focused beam[6], a time-intensive process that can lead to localised radiation damage and positional drift. Moreover, the point-scanning requirement of conventional XRF-CT makes direct integration with full-field PCT acquisition technically challenging, since the two modalities rely on fundamentally different acquisition geometries. Achieving in-situ and operando acquisition of both modalities would allow structural and chemical changes to be correlated under matched experimental conditions, offering a more comprehensive understanding of phenomena such as chemical reactions, phase transitions, and morphological evolutions. Currently, these techniques are performed sequentially, often on different instruments, which limits temporal resolution and makes it difficult to observe the same event across both modalities. Additionally, implementing XRF-CT in conjunction with complex sample environments can be technically challenging and may introduce artefacts or uncertainties. In this regard, sample translation, as required for raster-scanning XRF-CT, is often constrained by the sample environment. Bringing together low-resolution elemental mapping by XRF-CT with higher-resolution (PCT) mapping offers many opportunities in correlative tomography[7,8].

Previous efforts at full-field illumination XRF imaging have variously deployed pinholes[9,10], Fresnel lenses[11], poly-capillary arrays[12], and ghost imaging[13] (GI). These techniques illuminate the entire field of view, reducing data-acquisition time and mitigating dose-related effects. However, these approaches have not yet led to an experimentally demonstrated, calibrated three-dimensional full-field XRF tomography workflow compatible with co-registered attenuation or phase-contrast tomography. In the specific case of pinholes, which is most relevant to our work, tomographic reconstruction requires a dedicated calibration technique for all geometric parameters and degrees of freedom. Throughout this paper, we will refer to full-field XRF-CT using a hyperspectral detector as "**Full-Field Fluorescence Computed Tomography**" (**F3CT**).

Here, we propose F3CT, a synchrotron-based hyperspectral full-field fluorescence tomography technique enabled by a calibrated pinhole geometry and designed to be compatible with transmission/phase-contrast tomography acquired sequentially in the same experimental configuration and illumination geometry. Energy-resolved detectors have previously enabled 'spectral imaging' in other radiation modalities, for example in neutron Bragg-edge transmission imaging[14], where per-pixel wavelength information provides materials-specific contrast, including later in-situ studies of microstructural evolution under load[15,16]. Related hyperspectral X-ray chemical mapping has also proven valuable for non-destructive compositional analysis in challenging sample geometries, such as moon rock displays[17]. Here, we achieve F3CT by utilising a 2D pixelated photon-counting hyperspectral detector, HEXITEC[18], which has the ability to provide element-based maps in geology[17] and biology[19], in a shielded-aperture setup (i.e., using an aperture/mask manufactured from an X-ray shield). A key innovation in the present work is a calibration procedure that enables the correct interpretation and exploitation of the setup geometry for the F3CT reconstruction. This enables hyperspectral three-dimensional fluorescence reconstructions that can be co-registered with phase-contrast tomography acquired under the same experimental geometry. This provides a practical route toward correlated structural and chemical tomography, and establishes the geometry required for future concurrent multimodal implementations.

To our knowledge, this is the first experimental demonstration of synchrotron-based full-field hyperspectral X-ray fluorescence computed tomography in 3D with sequential phase-contrast/tomographic acquisition under the same illumination geometry, enabled by a pinhole-camera arrangement with a pixelated energy-resolving 2D detector and a dedicated calibration/reconstruction workflow for the virtual cone-beam geometry. Prior full-field XRF approaches have largely remained in 2D or focused on alternative optics schemes rather than experimentally demonstrated correlative three-dimensional fluorescence tomography compatible with co-registered transmission/phase-contrast imaging acquired under the same experimental geometry.

We apply F3CT to two case-study specimens chosen to test chemically and structurally distinct imaging scenarios. In the geological specimen, F3CT reveals barium- and iron-sensitive fluorescence distributions within a granite–serpentinite core stack that are not directly accessible from phase-contrast tomography alone. In the biological specimen, F3CT maps the three-dimensional distribution of a silver stain within a zebrafish trunk section. Together, these examples demonstrate the ability of F3CT to provide chemically sensitive volumetric information that can be interpreted alongside high-resolution structural tomography.

# 2 Methods

## 2.1 Experimental Details

We performed the experiment on the ID19 beamline at ESRF – The European Synchrotron, a hard X-ray beamline dedicated to phase-contrast and absorption microtomography[20] with an energy range of 15 to 225 keV. Figure 1 shows a schematic of the experimental setup and the associated tomographic images. The source was the ID19 w150 wiggler operated with a gap of 70 mm. Here, we used a parallel, polychromatic X-ray beam with an average effective energy of 59.25 keV, obtained by inserting 0.35 mm of copper and 2.8 mm of aluminium filters. Each sample was mounted on a standard rotation stage and illuminated in its entirety while reducing the illumination of the sample surroundings and minimising fluorescence contributions from the surrounding sample environment and support stage. This was achieved using a box beam with horizontal and vertical dimensions of 7.6 mm and 7.2 mm, respectively.

The pixelated, spectral HEXITEC[18] detector (with a 250 µm pixel size) was positioned at 90° to the incoming beam on an off-centre air-cushion platform along an optical arm to collect X-ray fluorescence images of the sample. An aperture disc, 5 mm thick, 20 mm diameter, and containing a 200 µm pinhole aperture in its centre, was mounted in front of the hyperspectral detector, as shown in Figure 2. By changing the distance between the pinhole and the sample and/or the detector, we could adjust the magnification, field-of-view, and corresponding effective pixel size of the hyperspectral images. Figure 2 (c) and (d) show the different distances used during the tuning process to find the best field of view for the samples, which are then calibrated with the calibration sample (chrome sphere) as shown in Figure 2 (c).

The calibration provides accurate distances, which have also been confirmed with physical measurements. The object–pinhole distance was ~79.0 mm, and the pinhole–detector distance was ~137.2 mm, yielding a magnification of ~1.74 and an effective pixel size of ~144 µm. Fluorescence projections were acquired as a triggered series using the HEXITEC detector over 121 angular steps, over 360°, with integration times of 170 s per projection for the granite and serpentinite cores stack dataset and 150 s for the zebrafish dataset, corresponding to approximately $10^6$ acquired frames per projection. The detector was operated without subsampling and without constraints on charge cluster size, and the raw frame data were reprocessed for spectral reconstruction and analysis.

For phase-contrast tomography, an indirect conversion (in-line) detector system was used, consisting of a 200 µm-thick LuAG:Ce scintillator coupled to an optical system with a ×2.1 magnification (Hasselblad), and a sCMOS-based pco.edge 5.5 camera (Excelitas, former PCO AG) with a native sensor resolution of 2560 × 2160 pixels, each 6.5 µm in size. The camera was positioned 450 mm downstream of the sample, which itself was located 145 m from the source. With the given magnification, the effective pixel size was 3.1 µm. A total of 5000 projections were acquired over a full 360° rotation.

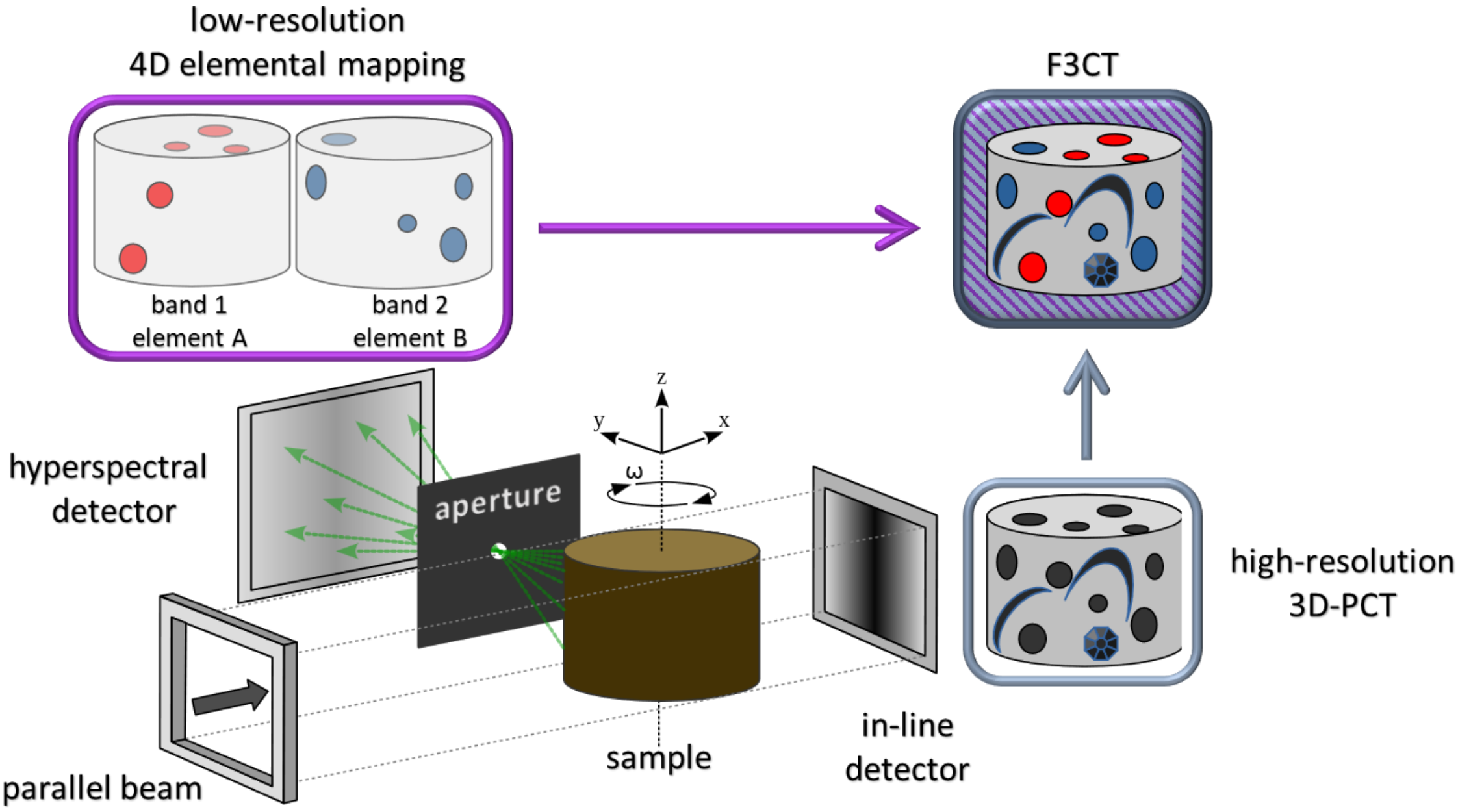

**Figure 1: Schematic of the F3CT experimental geometry. The sample is illuminated by a full-field synchrotron beam, fluorescence is encoded through a pinhole aperture and recorded by a hyperspectral 2D detector, while the transmitted X-rays are recorded downstream for phase-contrast tomography.**

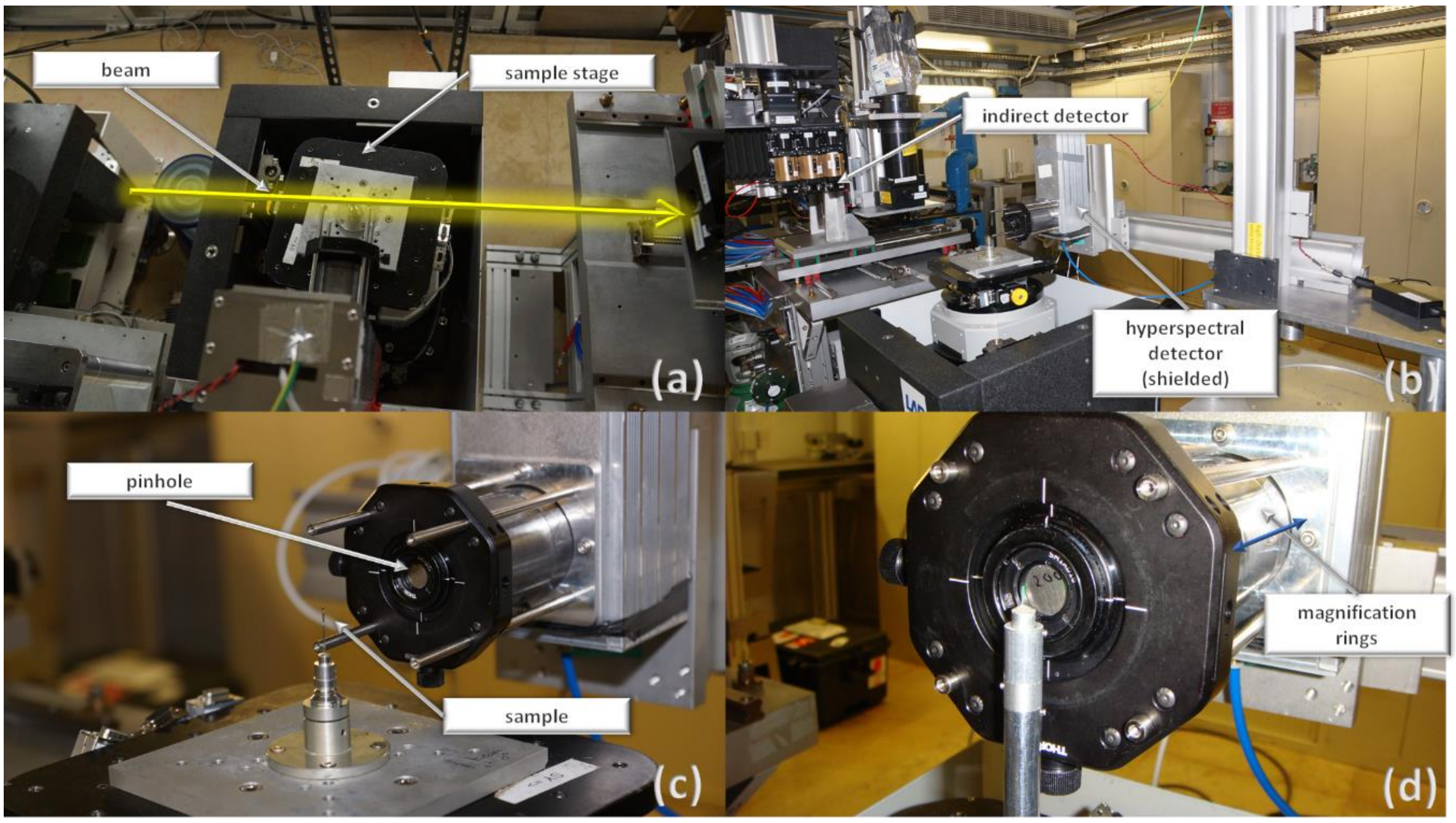

**Figure 2: Experimental implementation of the F3CT geometry. (a) Top view and (b) side view of the detector and sample arrangement, (c) sample stage region, and (d) close-up of the pinhole aperture.**

## 2.2 Calibration and Volume Reconstruction

Although the sample was illuminated by a parallel pink synchrotron X-ray beam, the pinhole detection geometry is mathematically equivalent to a cone-beam imaging geometry. This is presented in Figure 3, where the pinhole acts as a point source in front of the hyperspectral detector. The key difference from a conventional cone-beam system is that the object is represented at a virtual position relative to the pinhole and detector. In practice, this can be treated through a point-symmetry transformation of either the sample or detector coordinates.

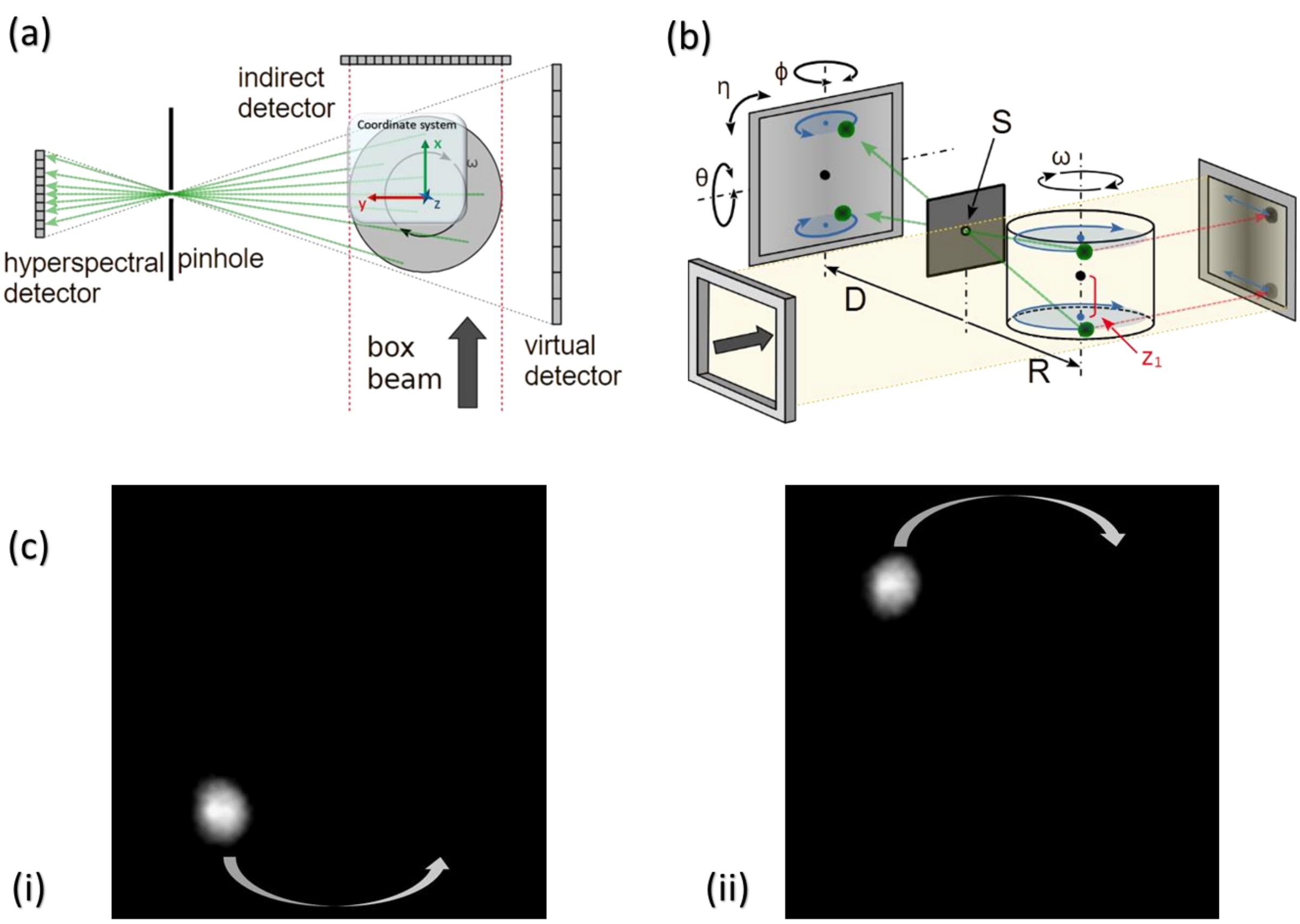

**Figure 3: Virtual cone-beam geometry used for F3CT reconstruction. (a) Plan view showing the pinhole, detector, and virtual detector representation; (b) three-dimensional geometry with the fitted degrees of freedom; and (c) calibration bead projections at two vertical positions used to estimate the geometric parameters.**

We used a two-stage approach to calibrate our representation of this geometry. First, we used the calibration procedure described in Noo et al.[21] to obtain an approximate estimation of the degrees of freedom in the geometry. We assumed that the detector was in the virtual position. Its coordinates are thus flipped relative to the point source, leaving the sample's rotation direction unchanged and matching the inline PCT geometry. Second, we optimised the geometry by performing a self-consistent data-driven fitting of acquisition parameters.

### 2.2.1 Pre-calibration (p)

**Step (p1):** align the sample rotation axis with the Centre of Mass (CoM) of the sample projections. This is done with the in-line detector, as is routine on synchrotron beamlines. The coordinate system is defined in Figure 3. The centring of the sample's CoM is obtained by offsetting the sample by half of the lateral shift between the image at 0° and the flipped image collected at 180°. The same procedure is used to centre the sample along the x-axis by using images at 90 and 270°.

For the following steps, it is preferable to use a small spherical object. Assuming that the calibration sphere is positioned at the centre between the rotation axis and the central axis of the spectral camera, it is first brought outside of the rotation axis by a known horizontal offset $r$, then raised to a known height $z_1$ above the centre of the images. At least 6 images should be acquired at different rotation angles between 0° and 360°. In practice, we acquired a total of 120 images for this experiment to reduce the problems caused by faulty pixels in the detector. The calibration sample is then lowered to a position $z_2$ below the centre of the acquired images, where the angular scan is repeated. In these two scans, as seen in Fig. 3, the sphere projection on the HEXITEC detector follows two ellipsoidal trajectories. The orientation and deformation of these two ellipsoids are used to reconstruct the geometry parameters. This is obtained with the following steps.

**Step (p2):** extract the spherical object's average position over the detector at each rotation angle. This step is necessary to later fit the ellipsoid positions, orientations, and deformations. For this step, one can either fit the object’s projection CoM in each image or use cross-correlation against a sample object image.

**Step (p3):** compute the centre and parameters of these two ellipses via least-squares minimisation from the set of object positions.

**Step (p4):** correct the detector rotation about its normal. The line connecting the centre of the two ellipses ($\mathbf{c}_1$ and $\mathbf{c}_2$ for the upper and lower centre, respectively) indicates the direction of the rotation axis over the spectral detector. Thus, the angle η determines the rotation of the detector around its normal vector (where $c_u$ and $c_v$ are the horizontal and vertical coordinates over the detector for the given point c):

$$\eta = \arctan\left(\frac{c_{u,1} - c_{u,2}}{c_{v,1} - c_{v,2}}\right) \quad (1)$$

**Step (p5):** fit the remaining parameter. The eccentricity of the two ellipses, once renormalised by $\boldsymbol{z_1}$ and $\boldsymbol{z_2}$, determines the horizontal tilt of the detector (i.e., angle θ). Instead, the relative angle between their longest axes can be used to estimate the vertical tilt of the detector (i.e., angle ϕ). The equations for computing these values are large and complex, and beyond the scope of this article. We thus refer to Noo et al.[21] for their in-depth description and explanation, and we provide an implementation of those equations, which is available online[22].

### 2.2.2 Data-driven optimisation

After pre-calibration, the calibration projections were used to refine the geometric model by minimising the tomographic reconstruction residual of the reference object:

$$\hat{p} = \underset{p}{argmin}\{L_p\} = \underset{p}{argmin}\left\{\frac{1}{2}\left\|W_p x - y\right\|_2^2\right\} \quad (2)$$

where **x** is the object reconstruction, **y** the acquired calibration images, **W** the projection operator, and **p** is the vector of geometric parameters. Thus, $\mathbf{W_p}$ is a projection operator that depends on the fitted parameters, $\mathbf{L_p}$ is the loss function associated with the parameter vector **p**, and $\hat{\boldsymbol{p}}$ is the optimal solution. While this is a non-linear minimisation problem, the pre-calibration procedure provides an initial solution that is close enough to an optimal solution. The problem should become mathematically well-behaved in that region.

In this work, the refinement was performed using a non-iterative parameter sweep around the pre-calibrated solution. For a given parameter, different tomographic reconstructions of the test object are performed at slightly different values around the pre-calibrated one. The value that minimises $\mathbf{L_p}$ is selected. We then further fine-tuned the selection by selecting the two adjacent tested values of the $\mathbf{L_p}$ curve, and performed a local parabolic fit to find the vertex position.

Alternatively, thanks to the advent of automatic differentiation tools, it is possible to iteratively solve Eq. (2) in Python using existing non-linear optimisation algorithms[23,24]. Although automatic differentiation offers a route to joint optimisation of all geometry parameters, the pre-calibrated solution was already sufficiently close to the optimum that a simpler local refinement was adequate for the present experiment, and it required only one iteration per parameter to achieve the required calibration precision.

## 2.3 Spectral Fluorescence Processing

The calibration of the HEXITEC energy channels was performed using an Americium 241 source. Per-pixel energy calibration was performed by fitting the detector response to known emission lines from an Am-241 source using a linear calibration model. Because additional electronic noise was present in the experimental hutch, a secondary calibration refinement was performed using the silver signal from the sample. Data from the HEXITEC detector are acquired as photon-counting events, with the deposited charge used to determine photon energy. The data were calibrated into 769 spectral bands spanning 4-100 keV with a step size of 0.125 keV.

The signal for the silver sample was sufficient to clearly identify the silver $k_{\alpha1}$ and a $k_{\beta1}$ emission lines. For weaker signals, where the XRF lines would superpose, an Internal Average Relative Reflectance[25] algorithm was applied to the spectral data. It consists of a pixel-based correction with the mean of the entire spectrum. Additionally, a non-local meets global iterative approach[26] was applied to the noisier datasets.

A maximum noise fraction (MNF) transform was applied to the hyperspectral fluorescence projections to separate dominant signal-bearing components from noise-dominated spectral variation[27]. Unlike standard PCA, which orders components by variance alone, MNF orders the transformed components according to image quality or signal-to-noise content. The six leading MNF-derived principal components were retained for visualisation and component-guided interpretation of the fluorescence projections. Peak positions were estimated from local maxima in the calibrated spectra, and three-dimensional fluorescence volumes were reconstructed by integrating 20 spectral channels around selected fluorescence regions of interest.

## 2.4 Case study samples

Although we have tested a wide range of samples and materials, here we focus on two case study specimens: the trunk of a silver-stained zebrafish (Danio rerio) and a stack of granite and serpentinite cylinders. In fluorescence, silver exhibits $K_{\alpha_1}$ and $K_{\beta_1}$ emission lines at 22.16 keV and 24.94 keV, respectively, and these are the dominant sample-derived fluorescence features expected from the stained zebrafish. The sample was held in a plastic vial during the experiment, and a phase-contrast tomography scan was acquired sequentially.

Serpentinite is a metamorphic rock derived from the alteration of ultramafic precursor materials and is composed predominantly of serpentine minerals (antigorite, lizardite, and chrysotile), often with accessory magnetite. It is characterised by a magnesium-rich silicate composition, whereas granite represents a silica-rich, feldspar-dominated lithology typical of the upper continental crust.

To place the expected fluorescence response of these materials in context, representative elemental compositions for granite and serpentinite were compiled from standard geochemical references (Rudnick and Gao, 2003[28]; Huang et al., 2015[29]) and plotted against their corresponding characteristic $K_{\alpha_1}$ emission energies (Figure 4). Only elements with fluorescence energies at or above the Ti $K_{\alpha_1}$ line (~4.5 keV) are considered, consistent with the effective detection threshold of the HEXITEC detector.

This representation highlights that fluorescence from major rock-forming elements such as Mg, Si, and Al lies below the detector sensitivity, whereas the measurable signal is dominated by higher-energy transitions from elements such as Fe, Ti, Mn, and, where present, trace elements including Cr and Ba. Consequently, these elements are expected to provide the primary contrast in the reconstructed fluorescence volumes.

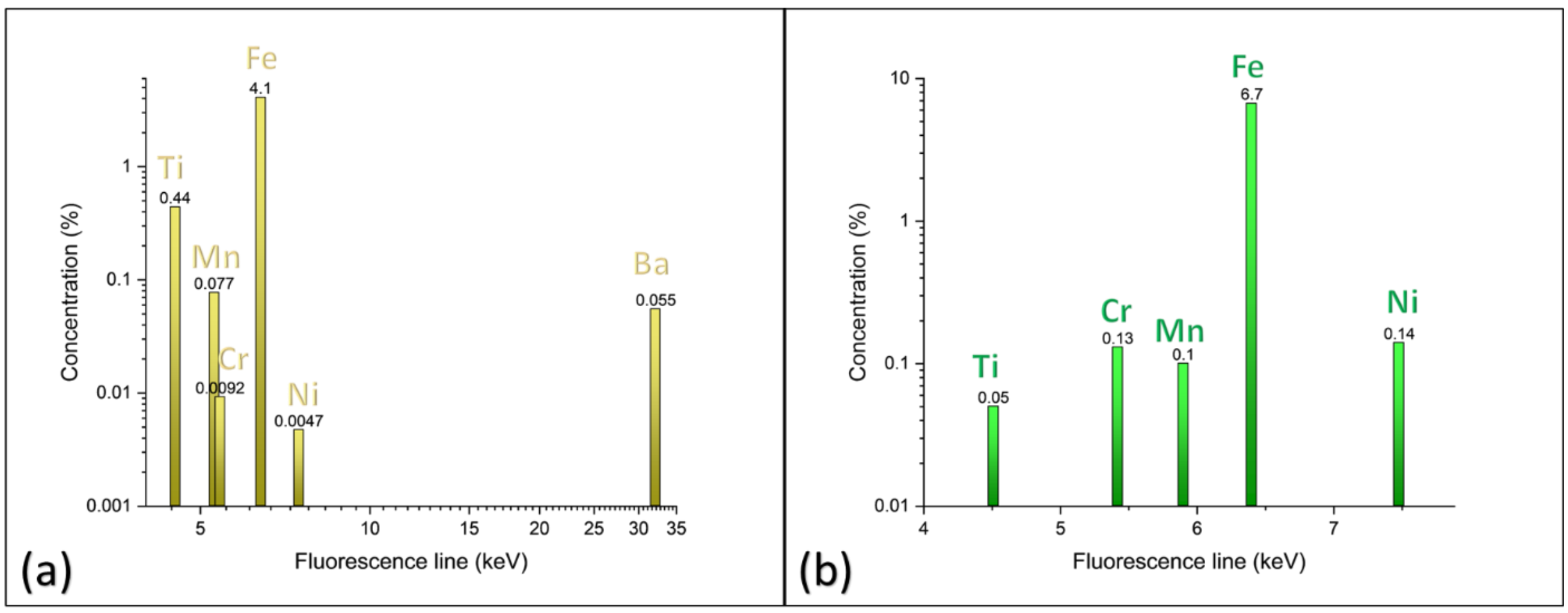

**Figure 4: Elemental composition and corresponding characteristic fluorescence energies for (a) granite and (b) serpentinite. Granite composition is based on Rudnick and Gao (2003), and serpentinite composition is derived from averaged serpentine measurements from Huang et al. (2015). Only elements with fluorescence energies ≥ Ti Kα (~4.5 keV) are shown. Panel (a) uses $\log_{10}$ scaling on both axes, while panel (b) uses $\log_{10}$ scaling on the concentration axis only.**

# 3 Results

## 3.1 Instrument performance

The F3CT system combines full-field illumination with hyperspectral fluorescence detection in a pinhole-camera geometry, enabling parallelised acquisition of elemental projection data without raster scanning. A total of ~120 fluorescence projections were acquired over 360°, compared with 5000 projections for the high-resolution phase-contrast tomography scan, highlighting the acquisition efficiency of the full-field approach.

The effective spatial resolution of the fluorescence reconstructions is on the order of ~150 µm, primarily governed by the pinhole aperture diameter, detector pixel pitch, and photon statistics. The hyperspectral detector enables separation of characteristic fluorescence lines within the recorded energy range, allowing identification of transition-metal and higher-Z emission features within a single acquisition. Despite reduced per-voxel photon statistics, the reconstructed fluorescence volumes retained sufficient signal-to-noise ratio to resolve elemental distributions in both biological and geological specimens.

Sensitivity to low-energy fluorescence emissions is limited by detector threshold effects and self-attenuation within the sample, resulting in a dominance of higher-energy elemental signals. Although the present implementation does not achieve the spatial resolution of focused pencil-beam XRF-CT, it operates in a complementary regime that prioritises acquisition efficiency, geometric simplicity, and compatibility with co-registered full-field multimodal tomography under matched experimental conditions.

## 3.2 Silver-stained zebrafish trunk

After phase-contrast tomographic reconstruction, the PCT volumes were segmented using a texture-classification machine-learning workflow based on seeded region propagation combined with watershed discrimination. The segmentation separated two dominant structural phases within the zebrafish trunk: the soft tissue, primarily composed of fibrous musculature (a), and the higher-attenuation dorsal vertebral bone (b), as shown in Figure 5. The reconstructed phase-contrast volume resolves the main anatomical features of the trunk and clearly reveals the spatial organisation of the musculature relative to the vertebral column. These high-resolution morphological data provide the structural framework required for subsequent correlation with the lower-resolution fluorescence volumes.

The hyperspectral fluorescence projections were analysed using the maximum noise fraction (MNF) transform described in section 2.3. The resulting component maps provide a compact representation of the dominant spectral–spatial contributions within the projection data and help identify regions contributing to the fluorescence signal.

The six leading MNF components (PC1–PC6) extracted from the zebrafish dataset are shown in Figure 6 for projection angle 126°, together with the corresponding spectrally colourised projection. Distinct experimental contributions can be identified within different components: PC1 is dominated by fluorescence from the Cu sample stage, PC4 from the plastic vial and surrounding medium, and PC3 is associated with the silver-stain contribution within the specimen.

Figure 7(a) presents the spectrum averaged over a central region of the projection, where the Ag $K_\alpha$ and Ag $K_\beta$ emission lines are clearly resolved at 22.16 keV and 24.94 keV, respectively. Additional intensity in the Cu/Zn fluorescence region is attributed to partial illumination of the metallic sample stage.

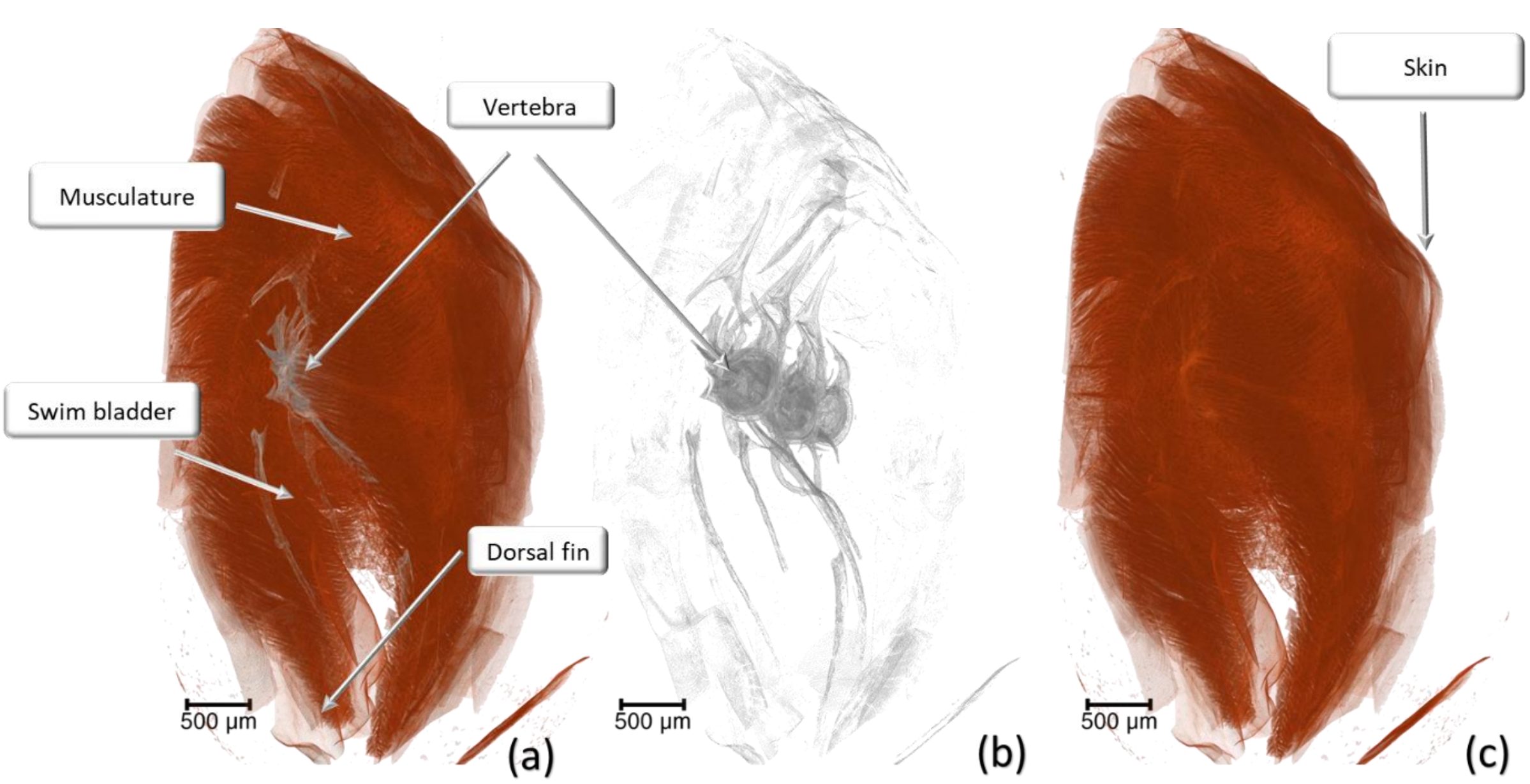

**Figure 5: Phase-contrast tomography (PCT) reconstruction of a transverse section of an adult zebrafish (Danio rerio) trunk. The images show a segmented 3D volume rendered to highlight anatomical structures. (a) Combined rendering showing soft tissue (brown) and skeletal structures (grey), with key anatomical features labelled. (b) Segmented rendering of the skeletal structure (vertebra) from the PCT volume. (c) Segmented rendering highlighting musculature and skin. All panels are displayed at the same scale and orientation.**

Three-dimensional fluorescence volumes were reconstructed by integrating spectral bands within a 20-channel window (2.5 keV) centred on the dominant fluorescence peaks. Figure 8 presents the resulting volumetric reconstructions. In the rendered overlay, the Ag $K_\alpha$ fluorescence contribution is shown in yellow, while the remaining integrated background signal, including contributions from the sample holder and surrounding medium, is displayed separately. The reconstructed fluorescence distribution demonstrates spatially varying silver-stain distribution throughout the soft tissue and confirms the ability of F3CT to recover spatially resolved elemental information in three dimensions under full-field illumination conditions.

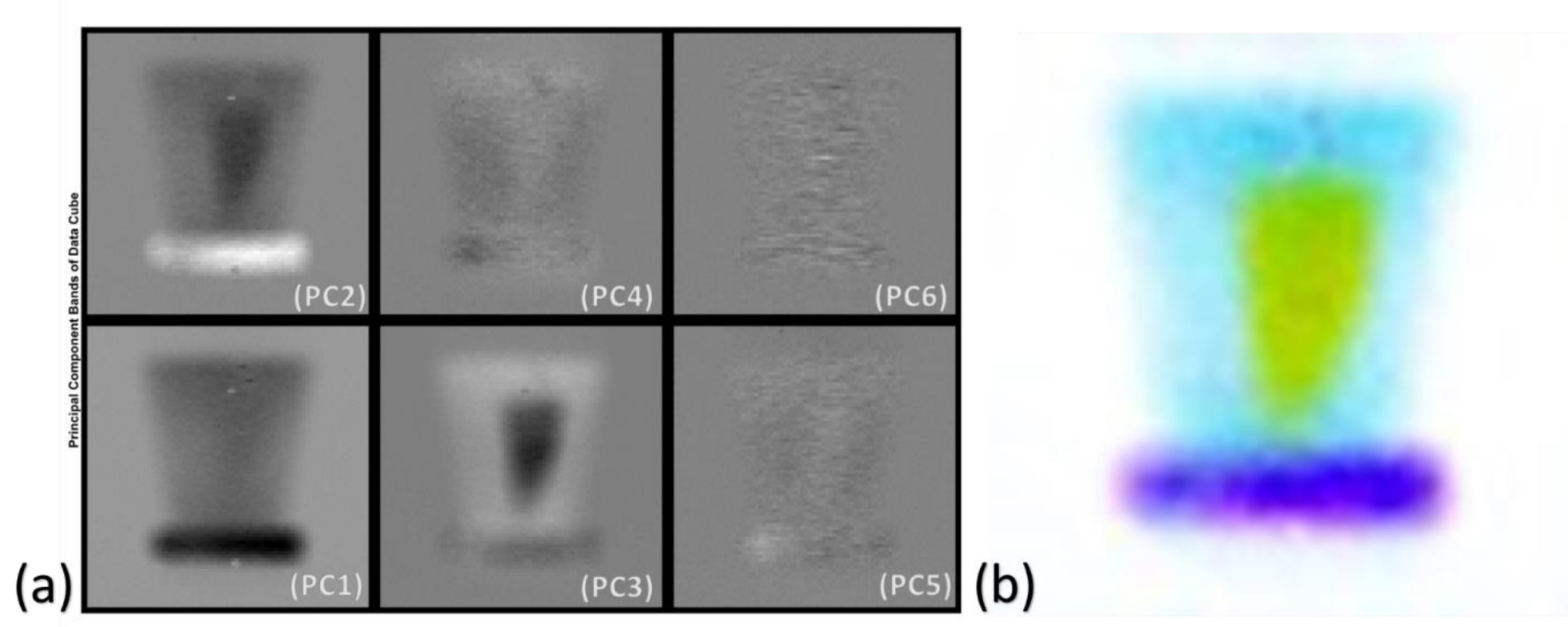

**Figure 6: (a) Six leading MNF-derived principal component maps extracted from the hyperspectral fluorescence projection of the silver-stained zebrafish trunk and (b) corresponding spectrally colourised projection at 126°. The component maps separate sample, holder, and background contributions within the full-field fluorescence projection.**

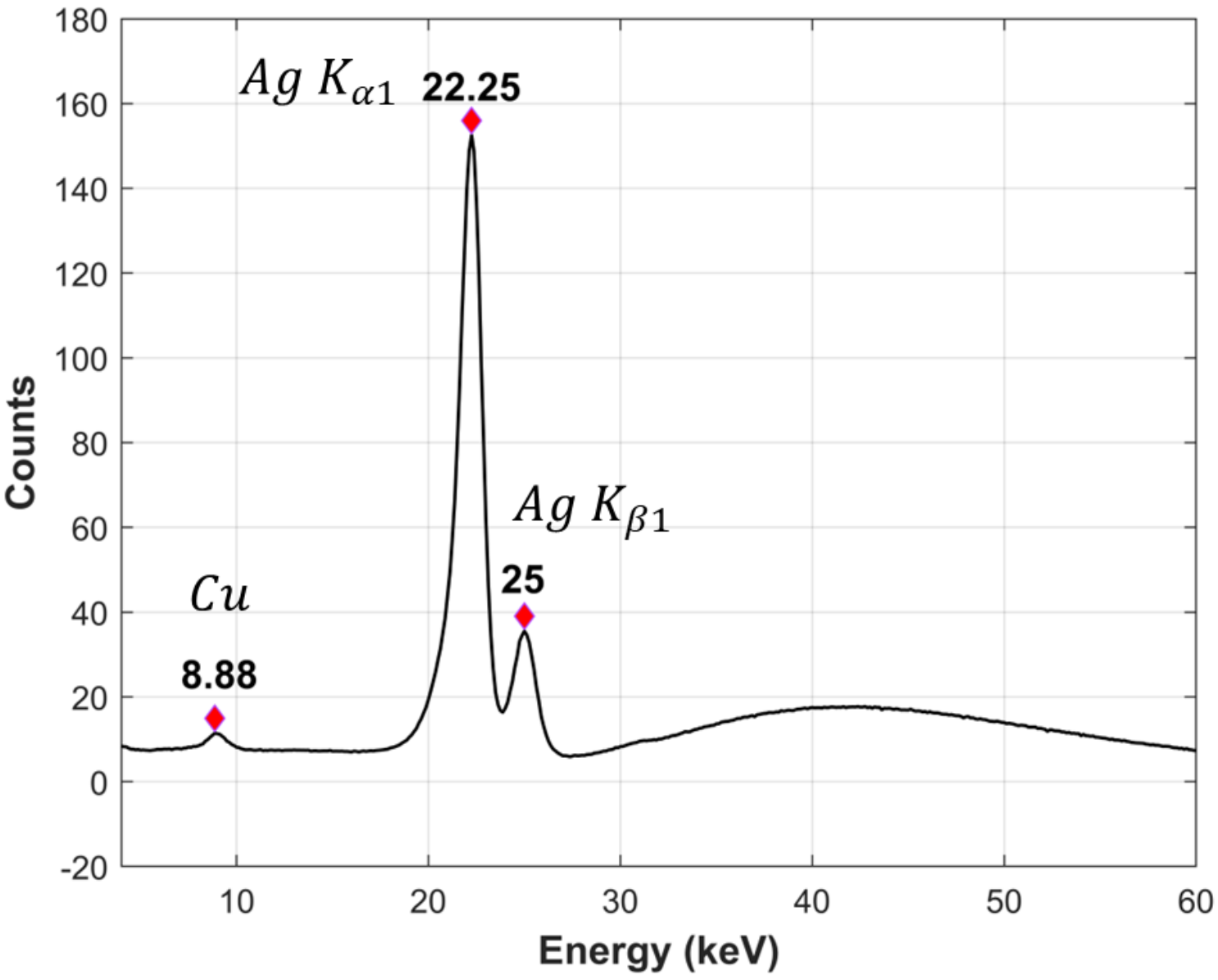

**Figure 7: Mean fluorescence spectrum acquired from the central region of a projection of the silver-stained zebrafish trunk at projection angle 126°. The Ag $K_\alpha$ and $K_\beta$ emission lines are clearly resolved at 22.16 keV and 24.94 keV. Additional intensity in the Cu/Zn region arises from partial illumination of the metallic sample stage.**

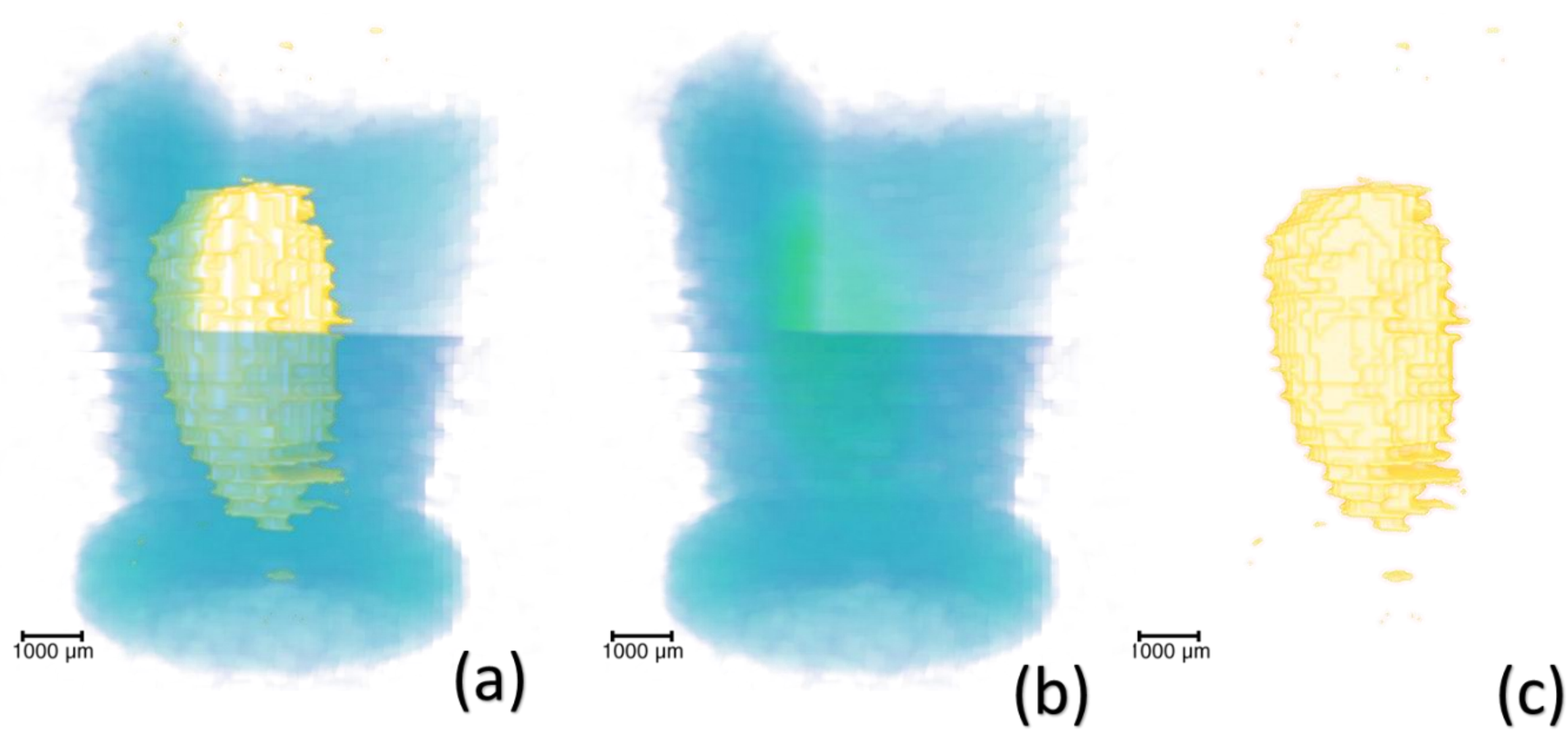

**Figure 8: Three-dimensional F3CT reconstruction of the silver-stained zebrafish trunk. (a) Combined fluorescence volume rendering showing both background contributions and the Ag-sensitive signal. A cubic cutout is applied to the volume to reveal internal features. (b) Background fluorescence contribution, shown with the same cutout to illustrate the surrounding medium and sample environment. (c) Ag-sensitive fluorescence signal, highlighting the spatial distribution of the silver stain within the specimen. The Ag signal is rendered in yellow/gold, while background contributions are shown in blue. All panels are displayed at the same scale and orientation.**

Overall, the zebrafish experiment demonstrates the ability of F3CT to recover chemically sensitive three-dimensional information from a biologically heterogeneous specimen while remaining compatible with high-resolution phase-contrast tomography acquired under the same experimental configuration. The phase-contrast reconstruction provides detailed structural context, whereas the hyperspectral fluorescence reconstruction maps the spatially varying distribution of the silver stain throughout the tissue volume. Although the fluorescence reconstruction remains lower in spatial resolution than the corresponding phase-contrast tomography, the experiment demonstrates that full-field hyperspectral fluorescence tomography can provide volumetric elemental mapping without raster-scanned excitation, while preserving compatibility with future in-situ and operando imaging workflows.

## 3.3 Granite and serpentinite core stacks

Geological materials are commonly investigated using non-destructive X-ray fluorescence methods for applications ranging from mineralogy, petrology, and resource geology to broader geoscience challenges such as fault-zone processes and subsurface $CO_2$ storage. However, fluorescence imaging of rocks remains challenging because many major rock-forming elements, including magnesium, silicon, sodium, and chlorine, emit predominantly low-energy fluorescence lines. These low-energy photons are strongly self-attenuated within the surrounding material and are additionally difficult to detect with CZT (cadmium zinc telluride) semiconductor material due to reduced detector sensitivity close to the low-energy threshold. In the present configuration, the HEXITEC detector exhibited a practical noise floor near 4 keV, limiting direct sensitivity to several light-element fluorescence bands. Consequently, the fluorescence signal is dominated by heavier or trace elements that emit at higher energies, which experience lower attenuation and propagate more efficiently through the sample volume.

The spatial distributions of the leading maximum noise fraction (MNF)-derived principal components extracted from the hyperspectral fluorescence dataset are shown in Figure 9(a), together with a spectrally colourised projection image in Figure 9(b) for projection angle 39°. The MNF transform was used as a signal-to-noise-ordered spectral decomposition of the multichannel fluorescence projections. In contrast to a simple energy-window image, which isolates signal over a predefined spectral range, the MNF decomposition identifies the dominant correlated spectral-spatial variations present across the full hyperspectral dataset while suppressing noise-dominated components. This makes it particularly useful for the granite–serpentinite specimen, where fluorescence from major low-Z rock-forming elements is weak or strongly attenuated, and where weaker sample-derived signals can be partially obscured by detector background, scattering, and fluorescence from neighbouring regions. In the colourised representation, selected MNF-derived PC maps are assigned to colour channels to visualise spatially distinct spectral contributions across the rock stack; in this rendering, PC1 is shown in green and PC3 in pink. The resulting image reveals clear differences between the granite and serpentinite regions, despite the comparatively weak fluorescence signal originating from the serpentinite. Several localised high-intensity inclusions are most evident in components associated with higher-energy spectral variation, while broader components capture more spatially distributed fluorescence contributions across the specimen. These component maps therefore provide a compact way to visualise chemically and spectrally heterogeneous regions prior to tomographic reconstruction.

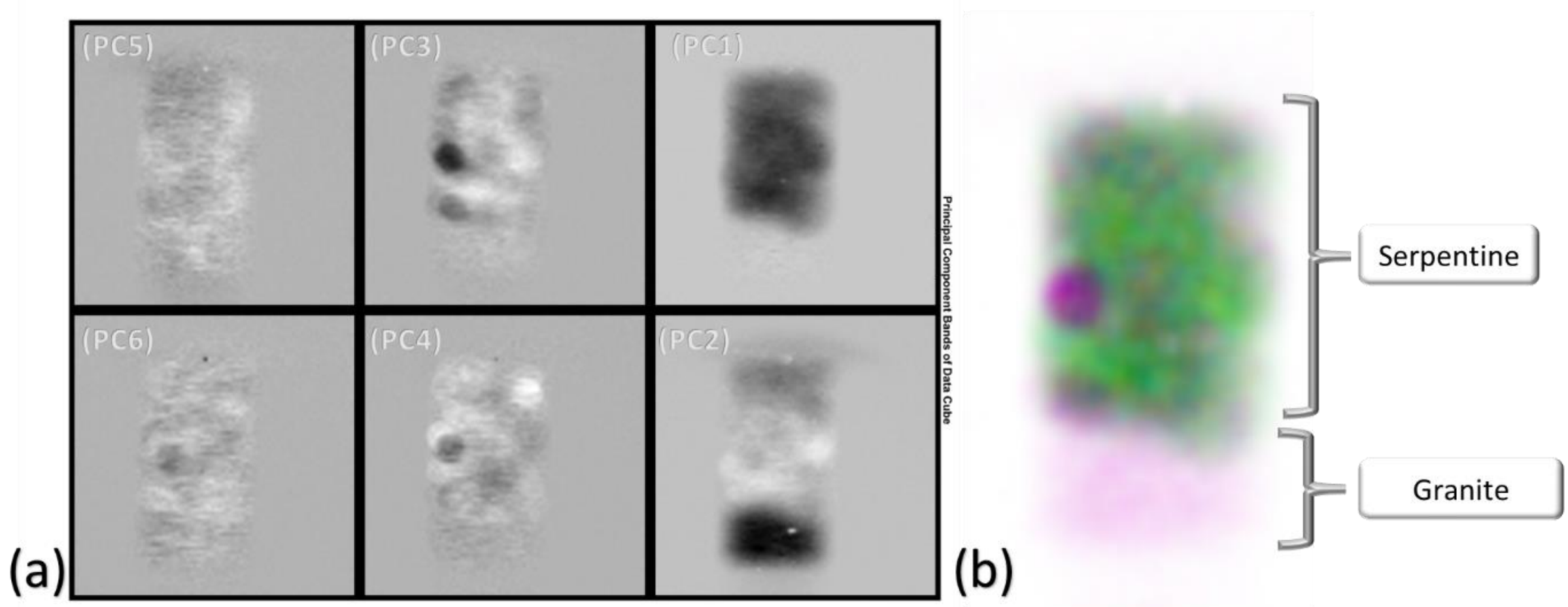

**Figure 9: Hyperspectral fluorescence analysis of the superimposed granite and serpentinite cylinders at projection angle 39°. (a) Spatial maps of the leading MNF-derived principal components and (b) spectrally colourised projection generated from selected component contributions. The colour channels represent component-derived spectral variations rather than direct elemental maps.**

It is important to note that, because MNF-derived principal components are noise-ranked decomposition modes rather than pure elemental maps, they should not be interpreted as direct maps of individual elements. The component features are therefore interpreted together with the mean fluorescence spectra and the integrated energy-window reconstructions, which provide the physically grounded link between spectral features and the reconstructed volumetric fluorescence signal.

Figure 10(a) presents the corresponding mean fluorescence spectrum acquired from the superimposed granite and serpentinite cylinders for the same projection angle. Several spectral features are visible, including a lower-energy transition-metal fluorescence region and prominent Ba-sensitive peaks in the 32–36 keV range. The associated MNF-derived principal components are shown in Figure 10(b). Note that

A feature near ~9.7 keV is consistent with tungsten $L_\beta$ fluorescence and is therefore attributed to beamline or experimental components (e.g. aperture disk or shielding), rather than intrinsic sample composition, although contributions from detector response effects cannot be fully excluded. A dominant peak near 6.4 keV with a higher-energy shoulder is observed and attributed to Fe $K_\alpha$ emission with a contributing $K_\beta$ component (~7.06 keV), broadened by the detector energy resolution and spectral binning; this is consistent with the presence of iron in both serpentinite and granite, where Fe is a common rock-forming element. Additional higher-energy features include Sr $K_\alpha$ (~14.2 keV) and Ba $K_\alpha/K_\beta$ (~32 and ~36 keV), which are consistent with trace elemental contributions, particularly in granite where incompatible elements such as Sr and Ba are commonly enriched. Peaks observed at ~23 and ~26 keV are assigned to Cd Kα and Kβ fluorescence originating from the CdTe detector material rather than the sample. A further feature near ~16 keV is observed but does not correspond precisely to expected Sr emission lines and is therefore not definitively assigned, potentially arising from trace high-Z elements (e.g. Nb or Zr) or detector-related effects. Note that the reported energies correspond to local maxima in the measured spectra rather than fitted peak positions, and small deviations from tabulated values are expected due to detector resolution, binning, and calibration uncertainties.

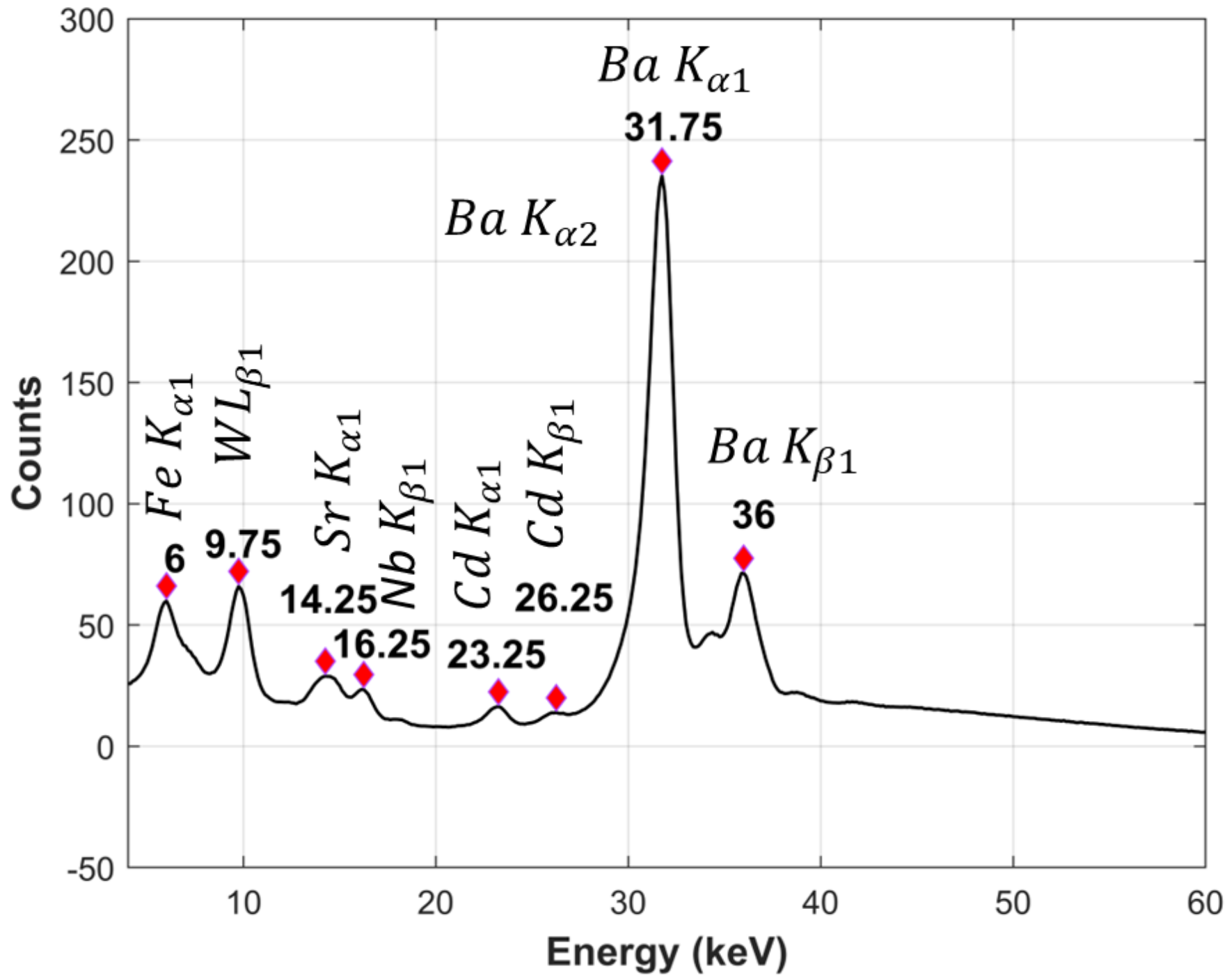

**Figure 10: (a) Mean fluorescence spectrum for the granite–serpentinite core stack acquired by F3CT at projection angle 39°. The spectrum contains lower-energy transition-metal fluorescence features and prominent Ba-sensitive peaks.**

The experiment was then extended to three-dimensional fluorescence tomography by acquiring a complete angular series of hyperspectral projections. The investigated specimen consisted of adjacent granite and serpentinite cylinders bonded together, as shown in Figure 11. This geometry was selected to test the separability of chemically distinct regions in a heterogeneous geological specimen and to assess the influence of scattered radiation on the reconstructed fluorescence signal. High-resolution phase-contrast tomography acquired under the same experimental configuration provided complementary structural information with an effective voxel size of 3.095 μm. The reconstructed phase-contrast volumes resolve structural features down to approximately 30 μm and enable identification of several mineralogical phases based on morphology and attenuation contrast. Within the serpentinite, magnetite-rich inclusions embedded within olivine-rich regions can be distinguished, while the granite volume exhibits characteristic microstructural features associated with quartz, plagioclase, microcline, and biotite-rich domains.

Given that several mineral phases are difficult to distinguish using attenuation contrast alone, the fluorescence reconstruction provides complementary chemical sensitivity. A clear distinction between the granite and serpentinite volumes is evident in the reconstructed fluorescence data, together with spatially localised inclusions associated with higher-energy fluorescence contributions. These features are most clearly observed in energy windows linked to the Ba-sensitive spectral region and are larger than many of the finer microstructural features visible in the phase-contrast segmentation. Broader spectral components capture more spatially distributed lower-energy fluorescence contributions across the specimen.

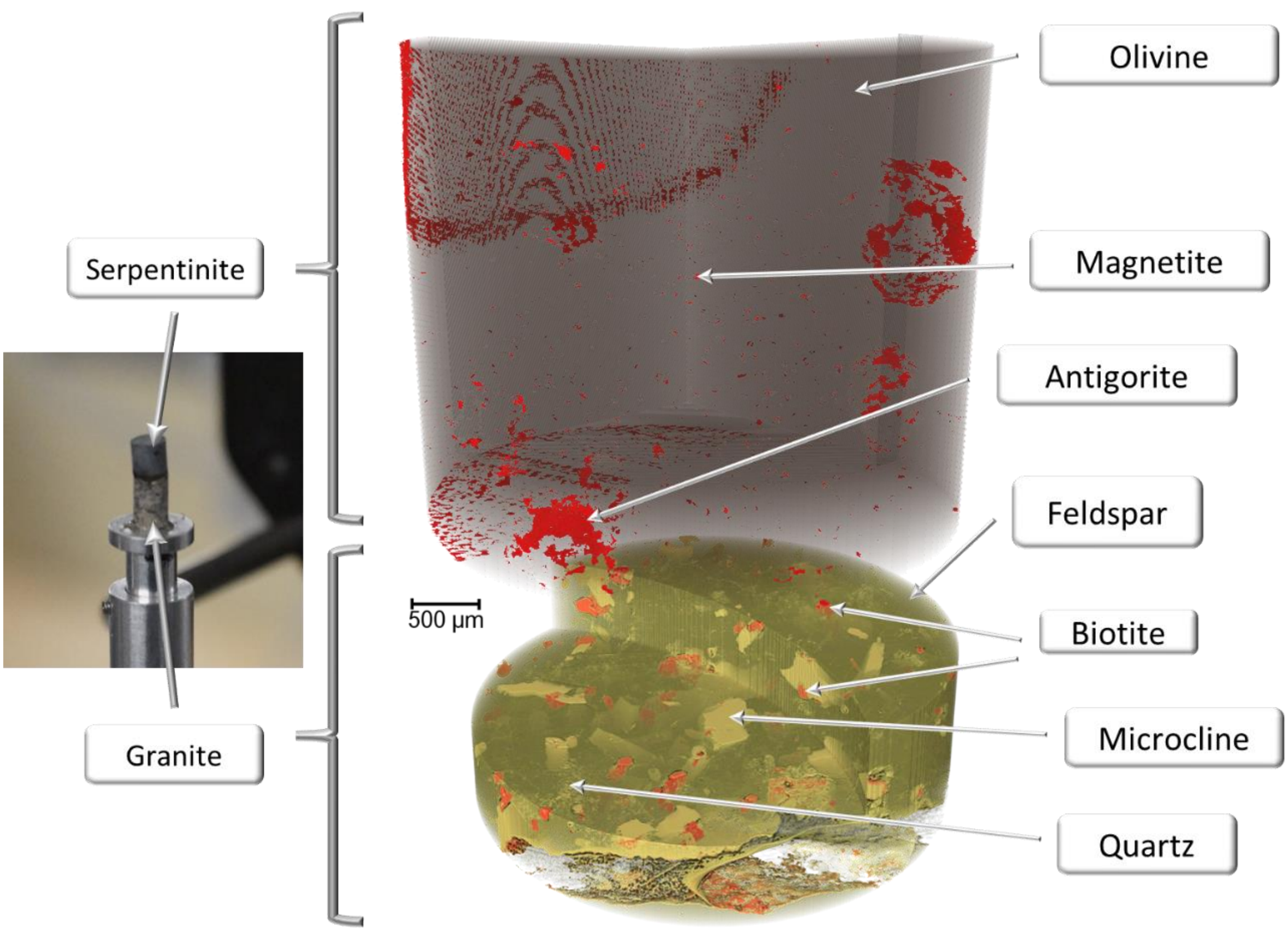

**Figure 11: Segmented synchrotron phase-contrast tomography reconstruction of the granite–serpentinite core stack. Colours are arbitrary and indicate segmented structural/mineralogical regions used as morphological context for comparison with the F3CT reconstruction.**

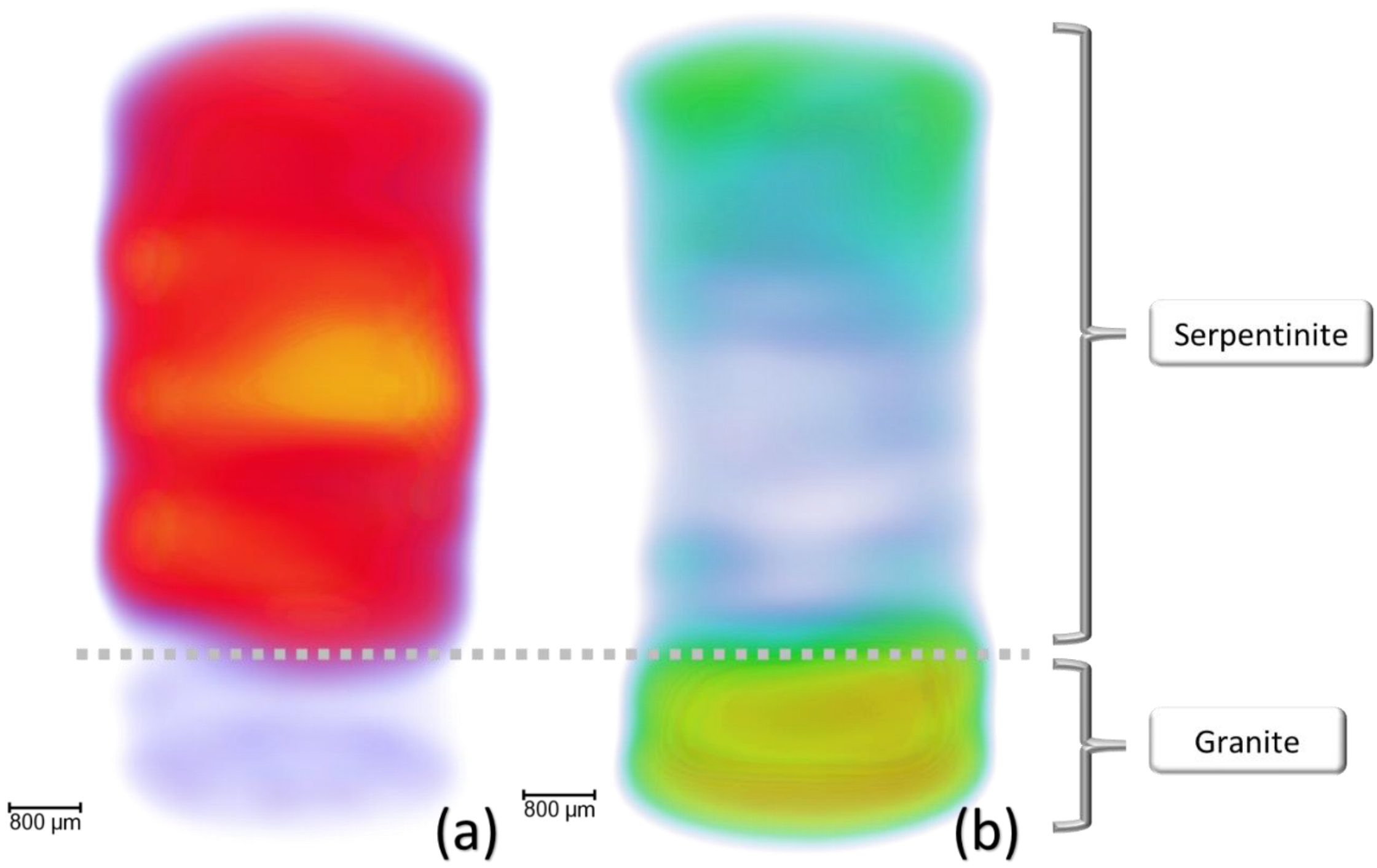

**Figure 12: Three-dimensional F3CT reconstructions of the granite–serpentinite core stack obtained from selected fluorescence energy windows. (a) Fe-sensitive fluorescence reconstruction, where the signal is shown in red with higher intensities rendered in bright orange. (b) Ba-sensitive fluorescence reconstruction, where the signal is shown in green with higher intensities rendered in yellow. Both panels are displayed at the same scale and orientation.**

Figure 12 presents three-dimensional fluorescence volumes reconstructed from selected energy windows. The reconstructions were obtained by integrating a narrow spectral window (20 channels) centred on the relevant fluorescence peaks. For the geological dataset, this approach provides a direct mapping of elemental contributions and is preferred over MNF-derived components. The Fe-sensitive and Ba-sensitive reconstructions highlight spatially distinct fluorescence contributions within the granite–serpentinite core stack, demonstrating the ability of F3CT to separate chemically distinct regions in a heterogeneous specimen.

The PCT reconstruction provides detailed morphological information on the granite–serpentinite stack, including the geometry of the bonded cores and the distribution of major structural/mineralogical regions. However, several phases within these rocks exhibit similar attenuation or phase-contrast signatures, making them difficult to distinguish reliably from morphology alone. The hyperspectral fluorescence reconstruction adds a complementary chemical channel by revealing spatial variations associated with higher-energy fluorescence features, including Ba-sensitive and Fe-sensitive contributions. These fluorescence signals highlight regions that are not directly identifiable in the PCT volume alone and therefore provide additional discrimination within the multiphase rock system. Importantly, the weaker fluorescence contribution from the serpentinite remains detectable despite the limited sensitivity to low-energy emissions from major light elements and the effects of self-attenuation within the sample. The combined PCT–F3CT datasets therefore demonstrate the value of the method for correlative mineralogical imaging: PCT supplies the structural framework, while F3CT adds chemically sensitive volumetric contrast without requiring raster-scanned excitation. This establishes F3CT as a useful approach for complex geological specimens where trace-element or high-energy fluorescence signatures are needed to complement conventional tomographic phase identification.

# 4 Discussion

## 4.1 Correlative PCT & F3CT imaging

The co-registered zebrafish data in Figure 13 illustrate the biological value of F3CT when interpreted alongside high-resolution phase-contrast tomography. The PCT volume provides the anatomical reference frame, resolving the overall morphology of the trunk, including the vertebral structure and surrounding soft tissues, while the F3CT volume reveals the three-dimensional distribution of the silver stain. In the overlay, the fluorescence signal is concentrated within selected internal soft-tissue regions rather than being uniformly distributed throughout the specimen. This comparison shows that the fluorescence reconstruction retains sufficient spatial fidelity to be interpreted directly against the higher-resolution tomographic anatomy, despite the difference in spatial resolution between the two modalities.

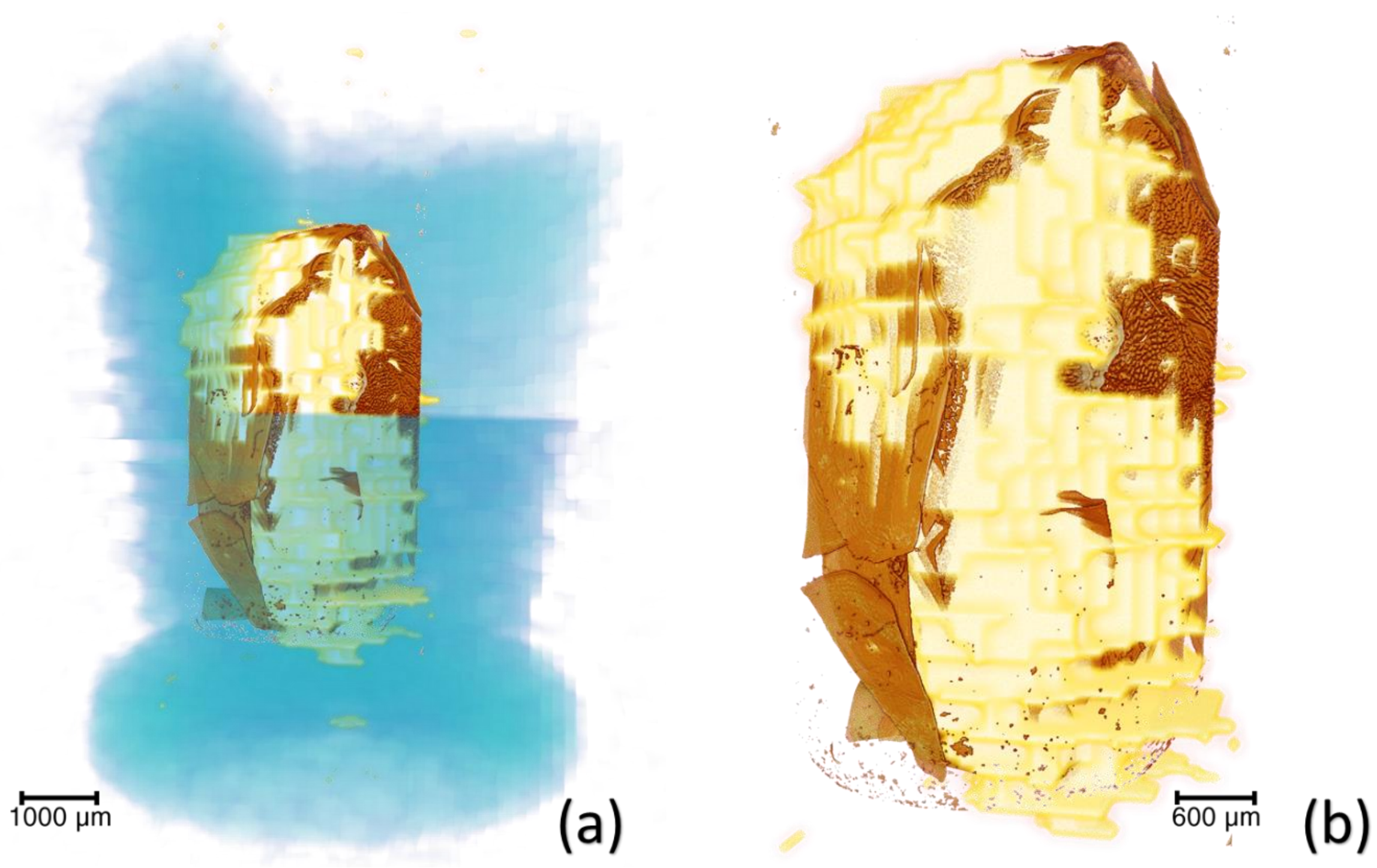

**Figure 13: Overlay of co-registered phase-contrast tomography and F3CT reconstructions for the silver-stained zebrafish trunk. The PCT volume provides anatomical context, while the F3CT reconstruction highlights the three-**

**dimensional distribution of the silver stain. Views are shown in the vial (a) and as a close-up of the specimen (b). The Ag-sensitive fluorescence volume is shown in gold/yellow, highlighting internal soft-tissue regions within the trunk.**

The geological case study presented in Figure 14 demonstrates that the combined phase-contrast tomography (PCT) and fluorescence tomography datasets clearly separate the granite and serpentinite components while simultaneously revealing chemically distinct regions within the reconstructed rock volumes. In conventional tomography, several silicate and feldspathic mineral phases exhibit very similar X-ray attenuation behaviour and therefore remain difficult to distinguish reliably using attenuation or phase contrast alone. In contrast, the fluorescence reconstruction introduces an additional chemically selective imaging channel that is sensitive to higher-energy elemental emissions. This allows regions with distinct spectral behaviour to be differentiated even when their structural contrast remains ambiguous. In particular, fluorescence-sensitive spectral regions associated with higher-energy elemental contributions reveal localised domains within the granite volume that cannot be directly inferred from morphology alone.

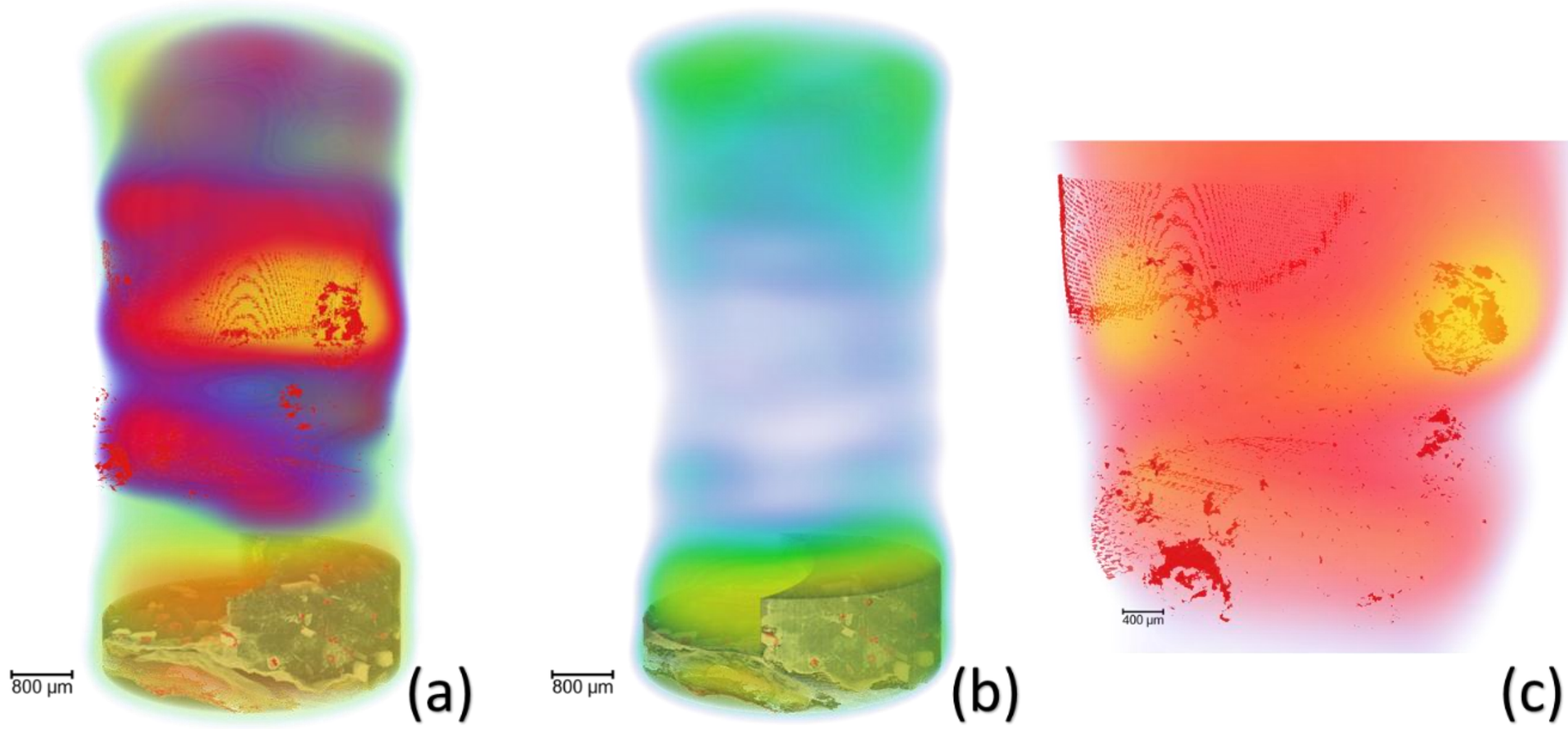

**Figure 14: Co-registered PCT and F3CT reconstructions of the granite–serpentinite core stack. (a) Fe-sensitive fluorescence overlaid on the PCT volume. (b) Ba-sensitive fluorescence overlaid on the PCT volume. (c) Zoomed serpentinite region showing the Fe-sensitive fluorescence relative to the PCT microstructure. Panels (a) and (b) are displayed at the same scale and orientation.**

The close-up view in Figure 14(c) illustrates how the Fe-sensitive fluorescence signal relates to the multiphase serpentinite microstructure. The fluorescence is not confined to discrete high-density inclusions, but also extends more diffusely throughout the surrounding regions of the core. When interpreted alongside the PCT reconstruction, the Fe-sensitive signal appears spatially heterogeneous within the broader serpentinite framework. This comparison highlights the value of co-registration: PCT provides the structural context, while F3CT reveals the spatial distribution of chemically sensitive features within that framework.

## 4.2 F3CT technique development

Three-dimensional X-ray fluorescence tomography has traditionally relied on raster-scanned pencil-beam (PB) acquisition, which is inherently sequential and time-consuming. On synchrotron beamlines, focused sub-micrometre beams can be achieved using dedicated optics[30], enabling high spatial resolution and elemental sensitivity, while laboratory implementations using cone-beam sources combined with pinhole geometries[31] remain similarly constrained by point-by-point acquisition. As a result, acquisition times remain long, particularly for large fields of view or high angular sampling. F3CT addresses this limitation by replacing raster-scanned point excitation with full-field illumination and spatially encoded hyperspectral detection. Each projection records fluorescence information across the entire field of view in a single acquisition, enabling tomographic reconstruction using a calibrated virtual cone-beam model. This approach provides a full-field formulation of X-ray fluorescence tomography and enables direct compatibility with co-registered phase-contrast tomography.

As with other imaging approaches exploiting HEXITEC-pinhole configurations[32,33], the present implementation involves a trade-off between throughput, spatial resolution, and spectral sensitivity. The spectral and spatial resolution remain lower than those achievable with raster-scanned systems employing high-purity germanium detectors, primarily due to charge-sharing effects and Compton scattering within the CZT sensor material. In its current configuration, the technique is therefore unlikely to rival the sub-micrometer spatial resolution capabilities achievable using state-of-the-art synchrotron pencil-beam systems. An additional practical limitation arises from the pinhole itself, since fabricating a self-supporting aperture with both sufficiently high X-ray attenuation and sub-25 μm feature size remains technically challenging. Consequently, the present implementation should be viewed as operating in a complementary regime that prioritises full-field acquisition, multimodal compatibility, and geometric simplicity over ultimate spatial resolution.

Several developments could substantially improve the performance of F3CT in future implementations. In particular, coded-aperture approaches could replace the single pinhole geometry in order to increase fluorescence throughput while preserving spatial separability through computational reconstruction[34]. As demonstrated in Soltau 2023[11] and Meyer 2025[35], this would increase fluorescence throughput while preserving spatial separability through computational reconstruction. Such approaches may additionally improve effective spatial resolution[36] when combined with dedicated deconvolution algorithms adapted to hyperspectral fluorescence datasets, as demonstrated for 2D images by the pan-sharpening approach of Meyer 2025[35]. On the detector side, the next generation of MHz HEXITEC detectors[37,38] is expected to provide substantial improvements in count-rate capability and temporal resolution, potentially increasing frame rates by several orders of magnitude. Together, these hardware and computational developments could enable significantly faster operando and in-situ fluorescence tomography experiments. Example applications include the investigation of dynamic processes in electrochemical energy-storage systems, such as tracking transition-metal migration and temperature evolution during Li-ion battery cycling[39,40]. The lower localised peak flux associated with full-field illumination may be relevant to future low-dose biomedical imaging strategies, including breast imaging for patients with high mammographic density, although such applications would require dedicated optimisation and validation[41,42].

The present implementation also highlights several remaining technical challenges. Current limitations include the trade-off between reduced flux and achievable spatial resolution imposed by the single-pinhole geometry, as well as the comparatively slow response of the current HEXITEC generation. The former limitation may be mitigated by using coded apertures with larger open-area fractions and spatially uncorrelated feature distributions, coupled with dedicated reconstruction and deconvolution strategies. The latter is expected to improve substantially with the introduction of next-generation high-frame-rate spectroscopic detectors. On the reconstruction side, additional improvements in geometry determination may also be achievable by incorporating tomographic consistency conditions into the calibration framework[43].

Full-field illumination changes the spatial and temporal distribution of incident flux relative to pencil-beam approaches. While a quantitative dose comparison is beyond the scope of the present work, the distributed illumination geometry may be advantageous for future studies of dose-sensitive specimens. This distinction is particularly important for dose-sensitive specimens, including biological tissues, where short-duration peak flux can induce substantially greater radiation damage than longer exposure to lower flux densities[44]. In addition, raster-scanned acquisition is inherently sensitive to sample drift and mechanical instability because each point in the reconstruction is acquired at a different time during the scan. In contrast, the full-field F3CT approach captures the entire field of view at each projection angle, enabling the application of established post-acquisition image registration and drift correction methods. This reduces the susceptibility of the reconstruction to motion-induced artefacts and further supports the suitability of the method for future time-resolved and operando imaging experiments.

# 5 Conclusions

We have proposed and demonstrated F3CT, a full-field hyperspectral fluorescence computed tomography technique based on a calibrated virtual cone-beam model and pinhole-encoded detection geometry. Using a pixelated photon-counting hyperspectral detector, F3CT reconstructs three-dimensional fluorescence volumes without raster-scanned excitation and remains compatible with phase-contrast tomography acquired under the same experimental geometry. To our knowledge, this is the first experimental demonstration of synchrotron-based full-field hyperspectral fluorescence tomography in three dimensions using a virtual cone-beam reconstruction framework.

The biological and geological case studies show how F3CT complements high-resolution structural tomography. In the zebrafish specimen, the method recovered the three-dimensional distribution of a silver stain throughout the tissue volume, while phase-contrast tomography provided the anatomical context. In the granite–serpentinite core stack, the fluorescence reconstruction revealed spatially varying high-energy spectral contributions that were not readily distinguishable from attenuation contrast alone. These results establish F3CT as a chemically sensitive volumetric imaging approach for heterogeneous specimens where structural contrast alone is insufficient.

More broadly, F3CT provides a full-field formulation of X-ray fluorescence tomography that complements high-resolution structural imaging. Although the current implementation does not match the spatial resolution of state-of-the-art pencil-beam XRF-CT, it operates in a regime defined by parallelised acquisition, simpler geometry, and direct compatibility with phase-contrast tomography. The method developed here therefore establishes an efficient complementary approach to correlative phase-contrast imaging, providing chemically sensitive volumetric information within the same experimental geometry.

## Acknowledgments

We are grateful for the proof-of-concept beamtime allocated to this project at ESRF ID19. We thank Frédéric Van Assche (Ghent University) for assistance with the initial calibration of the SpeXIDAQ software during beamtime setup.

We are particularly grateful to Sophie Sanchez and Tatjana Haitina (Evolutionary Biology Centre, Uppsala University) for providing the zebrafish sample used in this study, introduced via Paul Tafforeau (ESRF).

YC and PJW acknowledge funding from the European Research Council (ERC Grant No. 695638, CORREL-CT). We also acknowledge support from EPSRC grants “A Reconstruction Toolkit for Multichannel CT” (EP/P02226X/1) and “CCPi: Collaborative Computational Project in Tomographic Imaging” (EP/M022498/1 and EP/T026677/1). The Manchester (Henry Moseley) X-ray Imaging Facility is funded in part by EPSRC (Grants EP/F007906/1, EP/F001452/1 and EP/M010619/1) and forms part of the National Research Facility for laboratory X-ray CT (NXCT), supported by EPSRC (EP/T02593X/1).

This work was also supported by an EPSRC International Centre-to-Centre grant with the ESRF (EP/W003333/1), and PJW acknowledges the ESRF for hosting part of his sabbatical.

The supply of the HEXITEC camera system was funded by the UKRI STFC Centre for Instrumentation 2022–2023.

Beamtime was provided through ESRF in-house proposal IH-MI-1463.

## Ethical statement

All animal experimental procedures were approved by the local ethics committee for animal research in Uppsala, Sweden (permit numbers C161/4 and 5.8.18-18096/2019). The procedures were performed in accordance with the animal welfare guidelines of the Swedish National Board for Laboratory Animals.

## Silver staining

An adult zebrafish Danio rerio was euthanised with an overdose of MS-222 (300 mg/l), fixed in 4% paraformaldehyde solution in phosphate-buffered saline (PBS) and dehydrated in 25%; 50%, 75% ethanol series in PBS. A 5 mm thick transverse section of a trunk was cut and stained with silver ions for 2 hours. This was a control specimen for the unspecific staining in the immunostaining protocol with EnzMet (Nanoprobes, USA). Silver stained all the soft tissues (skin, muscles, and abdominal organs).

## Author Contributions

Thomas Zillhardt: **Conceptualisation, Investigation, Writing - Original Draft, Formal analysis, Methodology, Software, Data Curation, Validation, Visualisation**.

Yunhui Chen: **Conceptualisation, Investigation, Resources, Project administration, Methodology, Writing - Review & Editing**.

Alexander Rack: **Investigation**, **Data Curation, Methodology, Writing - Review & Editing**.

Matthew Veale: **Writing - Review & Editing**.

Matt Wilson: **Writing - Review & Editing.**

Philip J. Withers: **Conceptualisation, Methodology, Funding acquisition, Supervision, Writing - Review & Editing**.

Nicola Vigano: **Conceptualisation, Writing - Original Draft, Methodology, Software**.

## Conflicts of interest

The authors have no conflicts of interest to declare. All co-authors have seen and agree with the contents of the manuscript and there is no financial interest to report. We certify that the submission is original work and is not under review at any other publication.